\documentclass[aps,prl,reprint,floatfix,amsmath,amssymb,
superscriptaddress,footinbib]{revtex4-1}

\usepackage{graphicx}
\usepackage{epstopdf}
\usepackage{color}
\usepackage{bm}
\usepackage{printlen}
\usepackage{units}
\usepackage{dcolumn}

\usepackage{floatrow}

\usepackage[colorlinks=true,citecolor=cyan,linkcolor=magenta]{hyperref}

\usepackage{ulem}

\renewcommand{\vec}[1]{\mathbf{#1}}

\newcommand{\no}{\noindent}

\usepackage{nicefrac}

\definecolor{darkgreen}{rgb}{0.0, 0.26, 0.15}

\definecolor{rred}{rgb}{0.77, 0.12, 0.23}

\definecolor{orange}{rgb}{1.0, 0.49, 0.0}

\begin{document}

\title{Moir{\'e} Flat Bands in Twisted Double Bilayer Graphene}

\author{Fatemeh Haddadi}
\affiliation{Institute of Physics, Ecole Polytechnique F\'{e}d\'{e}rale de Lausanne (EPFL), CH-1015 Lausanne, Switzerland}

\author{QuanSheng Wu}
\email{quansheng.wu@epfl.ch}
\affiliation{Institute of Physics, Ecole Polytechnique F\'{e}d\'{e}rale de Lausanne (EPFL), CH-1015 Lausanne, Switzerland}
\affiliation{National Centre for Computational Design and Discovery of Novel Materials MARVEL, Ecole Polytechnique F\'{e}d\'{e}rale de Lausanne (EPFL), CH-1015 Lausanne, Switzerland}

\author{Alex J. Kruchkov}
\email{akruchkov@g.harvard.edu}
\affiliation{Department of Physics, Harvard University, Cambridge, MA 02138, USA}

\author{Oleg V. Yazyev} 	
\email{oleg.yazyev@epfl.ch}
\affiliation{Institute of Physics, Ecole Polytechnique F\'{e}d\'{e}rale de Lausanne (EPFL), CH-1015 Lausanne, Switzerland}
\affiliation{National Centre for Computational Design and Discovery of Novel Materials MARVEL, Ecole Polytechnique F\'{e}d\'{e}rale de Lausanne (EPFL), CH-1015 Lausanne, Switzerland}

\date{\today}

\begin{abstract}
We investigate twisted double bilayer graphene (TDBG), a four-layer system composed of two AB-stacked graphene bilayers rotated with respect to each other by a small angle. 
Our \textit{ab initio} band structure calculations reveal a considerable energy gap at the charge point neutrality that we assign to the \textit{intrinsic symmetric polarization} (ISP). 
We then introduce the ISP effect into the tight-binding parameterization and perform calculations on TDBG models that include lattice relaxation effects down to very small twist angles.
We identify a narrow region around the magic angle $\theta^\circ = 1.3^{\circ}$ characterized by a manifold of remarkably flat bands gapped out from other states even without external electric fields. 
To understand the fundamental origin of the magic angle in TDBG, we construct a continuum model that points to a hidden mathematical link to the twisted bilayer graphene (TBG) model, thus indicating that the band flattening  is a fundamental feature of TDBG, and is not a result of external fields.    
\end{abstract}

\maketitle

Twisted bilayer graphene (TBG) has recently attracted   considerable attention following the
discovery of the correlated insulator and superconducting phases  \cite{Cao2018a,Cao2018b, Yankowitz2018}, if tuned to the so-called \textit{magic angle} of 1.1$^\circ$, at which almost dispersionless (flat) electronic bands emerge near the charge neutrality point \cite{Laissardiere2010, Bistritzer2010,Guinea2012,Gargiulo2017,magic1,Carr2019}. 
The appearance of flat bands at magic angles in TBG is not just a coincidence in material properties engineering, but a fundamental feature of the TBG-like Hamiltonians \cite{magic1,magic2}. In such models, the flat band in the electronic spectrum appears if two Dirac cones are brought sufficiently close in reciprocal space via the moir{\'e} interlayer potential, so they hybridize \cite{Santos2007} and degenerate into remarkably flat bands \cite{Bistritzer2010,magic1}. In this regard, one can consider an extension of this picture to the hybridization of touching (gapless) pairs of parabolic bands that are primarily flatter at the touching point. Such scenario can be realized, for example, in the Bernal-stacked (AB-stacked) bilayer graphene characterized by touching parabolic bands at the $K$ and $K'$ points in the Brillouin zone \cite{CastroNeto2009}.  This idea has proved to be fruitful as witnessed by the recent experimental reports of unconventional superconductivity and spin-polarized insulating phases in twisted double bilayer graphene (TDBG) at twist angle $1.24^{\circ}$ \cite{Liu2019,Cao2019,Shen2019,Lee2019}.   
Given the current progress in  manufacturing TDBG, and strong sensitivity of the superconducting phase to the factors that modify the band structure (electric displacement and in-plane magnetic fields), one of the key questions is understanding the very nature of band flattening in TDBG and the evolution of the band structure upon changing the twist angle.

\begin{figure}[b]
\vspace{-10 mm}
\includegraphics[width=1.00 \columnwidth]{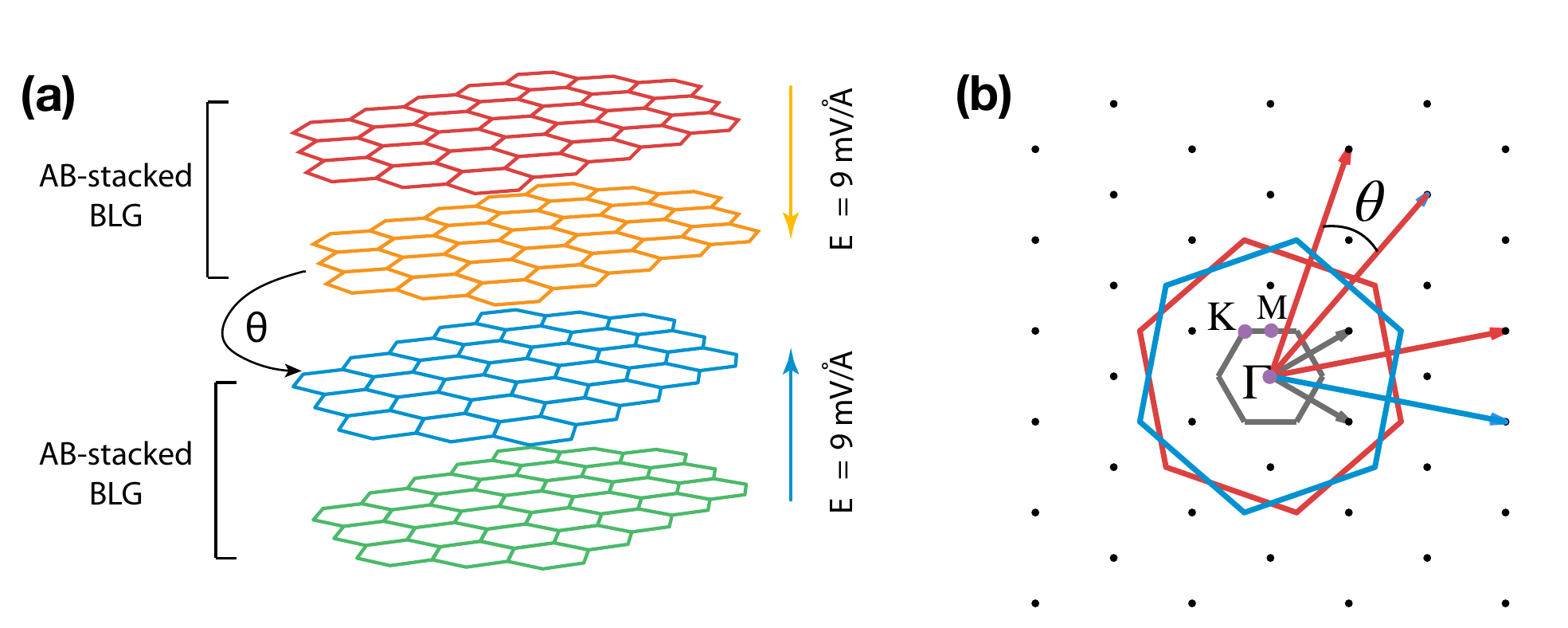}
\caption{ (a) Schematic drawing of the twisted double bilayer graphene configuration investigated in our work. The intrinsic symmetric polarization (ISP) is illustrated. (b) Reciprocal lattice (black dots) and the mini Brillouin zone (grey hexagon) of $\theta = 21.8^\circ$ ($m=1$) TDBG. The blue and red hexagons are the BZs the corresponding AB-stacked BLG counterparts.}
\label{fig1}
\end{figure}

In this Letter, we address the moir{\'e} flat bands in {TDBG} using \textit{ab initio} and tight-binding approximation calculations as well as an effective continuum model. Our DFT calculations performed on TDBG models with twist angles down to 2.5$^\circ$ reveal the presence of band gaps as large as 35~meV. These persistent band gaps are explained by the intrinsic polarization of  individual TBG components due to the proximity with the complementary TBG counterpart. By including this intrinsic symmetric polarization (ISP), an effect that has not been considered previously, into the tight-binging (TB) model we are able to perform accurate band structure calculations on TDBG models that include lattice relaxation effects down to very small twist angles.
In addition gap opening, the ISP flatten the electronic bands near the charge neutrality point. We also find a well-defined magic angle of $1.3^{\circ}$, at which both electron and hole gaps are maximized. Outside the narrow range of twist angles around $1.3^{\circ}$ the bands become highly dispersive, in contrast to the case of twisted transition metal dichalcogenides bilayers \cite{Wu2019,Naik2018} where the magic angle is hard to define. 
To understand the origin of magic angle in TDBG, we  further construct an effective continuum model describing the behavior at large moir{\'e} periodicities. This allows us to understand the mechanism of  band flattening in TDBG through the exact mapping onto the continuum TBG Hamiltonian \cite{magic1} if the particle-hole asymmetry is excluded.

\twocolumngrid

{\onecolumngrid
\begin{figure}[t]
\floatbox[{\capbeside\thisfloatsetup{capbesideposition={right,center},capbesidewidth=4.5cm}}]{figure}[\FBwidth]
{\caption{\label{fig3} {Band structures of (a) $\theta = 5.09^{\circ}$ and (b) $\theta = 2.45^{\circ}$ TDBG models calculated using DFT, the tight-binding model (TB) and the tight-binding model with the intrinsic polarization effect taken into account (TB+ISP).}}}
{\includegraphics[width=12 cm]{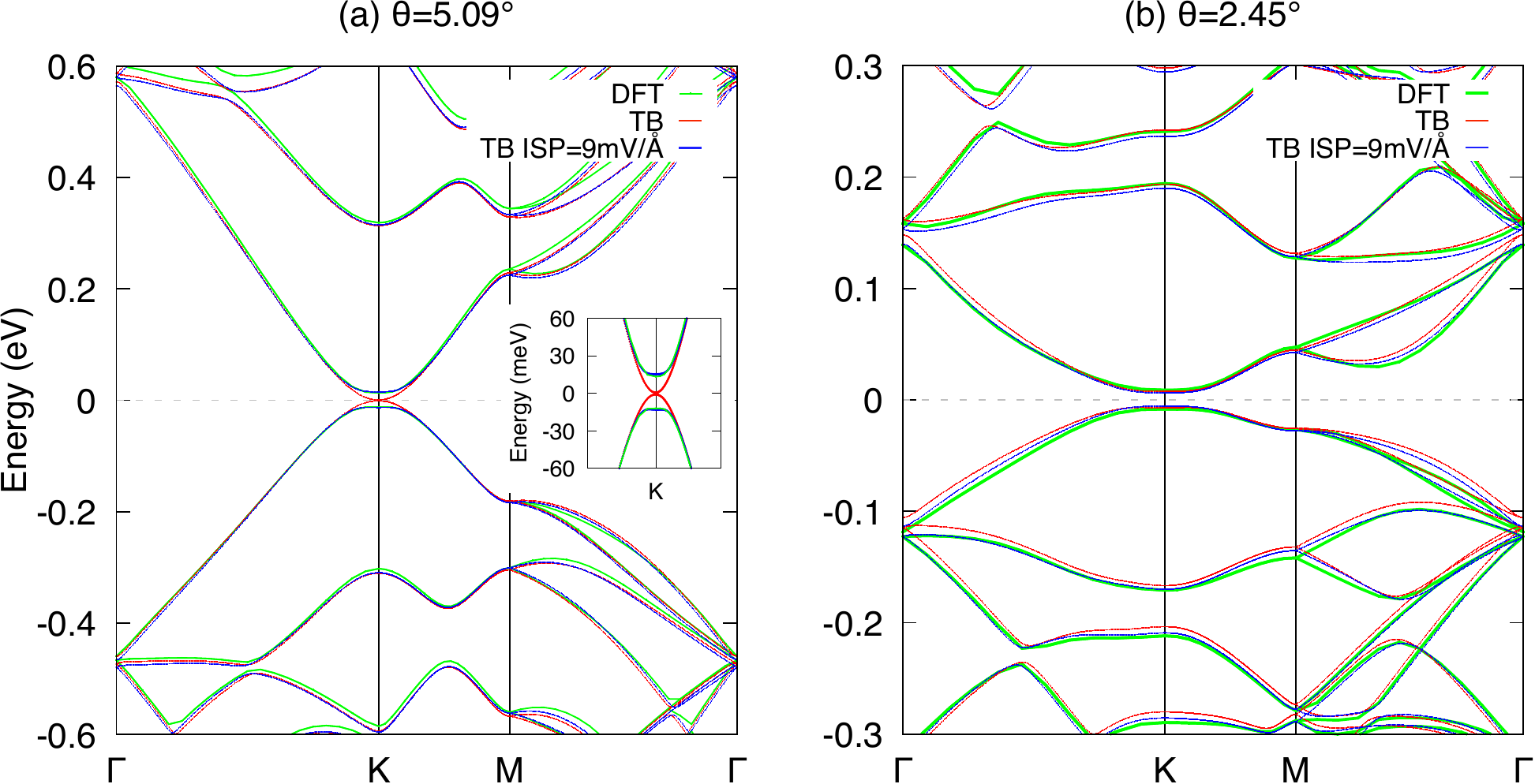}}
\end{figure}
\twocolumngrid
}

\twocolumngrid

Twisted double bilayer graphene (TDBG)  is a very special configuration among the large
family of graphene multilayers, which has gained a well-deserved attention due to the characteristic moir{\'e} physics (see e.g. Refs~\onlinecite{magic2,Cea2019,Mora2019,Zhang2019,Koshino2019,Choi2019}). 
 In the most general setting, the geometric overlay of four lattices leads to a sophisticated pattern of three interfering moir{\'e} superlattices, which may or may not be periodic and commensurate.   A more fruitful and simpler situation for both experiment and theory is the four-layer graphene configuration with only one twist parameter. This can be achieved in two ways, either twisting an AB-stacked bilayer on another AB-stacked bilayer by angle $\theta$, or working in the twist-alternate setting  \cite{magic2}. 
We focus on the former situation shown in Fig.~\ref{fig1}a that addresses recent experimental works \cite{Liu2019,Cao2019,Shen2019,Lee2019}.  The corresponding mini Brillouin zone  of the moir{\'e} superlattice (Fig.~\ref{fig1}b) is the same as in the case of TBG. 
Below, we consider TDBG models that correspond to the principal moir{\'e} branch with a discrete set of periods  $\lambda (\theta_m) = a/ 2 \sin (\theta_m/2)$, see Refs.~\onlinecite{Santos2007,Shallcross2008a,Shallcross2008b}.  
For every commensurate twist angle $\theta_m$, the supercell periodicity vectors are defined as
$ \mathbf{t_1} = m \mathbf{a_1} + (m+1) \mathbf{a_2} $ and $\mathbf{t_2} = -(m+1) \mathbf{a_1} + (2m + 1) \mathbf{a_2} $, where $a_{1,2} = a ( \sqrt{3} / 2 , \pm 1/2)$ are the lattice vectors of graphene with lattice constant $a = 2.46$~\AA\ and $m$ is commensuration condition parameter, 
$\cos \theta_m = (3m^{2} + 3m + {1}/{2})/(3m^{2} + 3m + 1)$.

\textit{DFT calculations.}---
We performed DFT calculations 
\footnote{Our DFT calculations have been performed using the Vienna \textit{ab initio} simulation package (VASP) \cite{Kresse} within the GGA approximation. The cutoff energy of $400$~eV was chosen for the plane wave basis. The Brillouin zone was sampled using meshes of $N \times N$ \textbf{k} points, where $N=36,24,12,9,6,6$  for the TDBG models described by $m=1,..,6$, respectively. The total energy convergence threshold was set to $10^{-7}$ eV. } 
on TDBG models characterized by commensuration parameters $m$ up to $m = 13$, which corresponds to $\theta = 2.45^\circ$ and includes 4376 atoms per supercell. Calculations on larger models for smaller twist angles  are unaccessible for computational cost reasons. 
Figure~\ref{fig3} shows the band structures of $m = 6$ ($\theta = 5.09^\circ$) and $m = 13$ ($\theta = 2.45^\circ$) TDBG models. One can immediately notice a gap opening where the two parabolic bands characteristic of AB-stacked BLG are supposed to touch. Table~\ref{t1} further confirms the systematic presence of few tens meV band gaps in $m=1,..,6$ TDBG models when DFT calculations are considered.
Interestingly, these band gaps are not qualitatively reproduced by the conventional tight-binding (TB) model \cite{Laissardiere2012} widely used for studying twisted multilayer graphene, even though gap opening in a certain range of twist angles has already been pointed out \cite{Choi2019}. This allows us to conclude that the mechanism responsible for this gap opening is not accounted for by the standard TB model. Indeed, each of the AB-stacked BLG counterparts in TDBG is placed in an asymmetric environment with one of the graphene layers facing vacuum and another facing adjacent graphene layer. This produces a slight potential difference between the two graphene layers which opens a band gap in otherwise gapless AB-stacked BLG. Below, we will refer to this effect as the intrinsic symmetric polarization (ISP). Fig.~\ref{fig1}a depicts it as effective electric fields that have opposite orientation in the two BLG subsystems. We find that introducing a single universal value of 9~mV/\AA\ for the electric field into the tight-binding model allows to reproduce the DFT results for all investigated TDBG models (TB+ISP in Table~\ref{t1} and Fig.~\ref{fig3}).

\begin{table}[b]
\caption{\label{t1}  Comparison of the band gaps (in meV) for TDBG models characterized by different commensurate parameters $m$ twist angles $\theta$ calculated using DFT, conventional tight-binding model (TB) and the TB model that includes the intrinsic symmetric polarization (TB+ISP). }
\centering
\begin{ruledtabular}
\begin{tabular}{l*{6}{d}r} 
$m$          & 1 & 2 & 3 & 4 & 5 &6 \\
$\theta$ ($^{\circ}$)   & 21.79   &  13.17 & 9.43 & 7.34 & 6.01 & 5.09 \\
\hline
DFT  & 30.08 & 34.41 & 31.45 & 29.88 & 23.03 & 25.26 \\
TB & 1.16  & 3.31  & 2.54  & 1.41  & 0.02  & 1.58  \\
TB+ISP   & 30.94 & 33.06 & 32.34 & 31.20 & 29.75 & 28.05
\end{tabular}
\end{ruledtabular}
\end{table}

\twocolumngrid

 \begin{figure*}
\includegraphics[width=1 \textwidth]{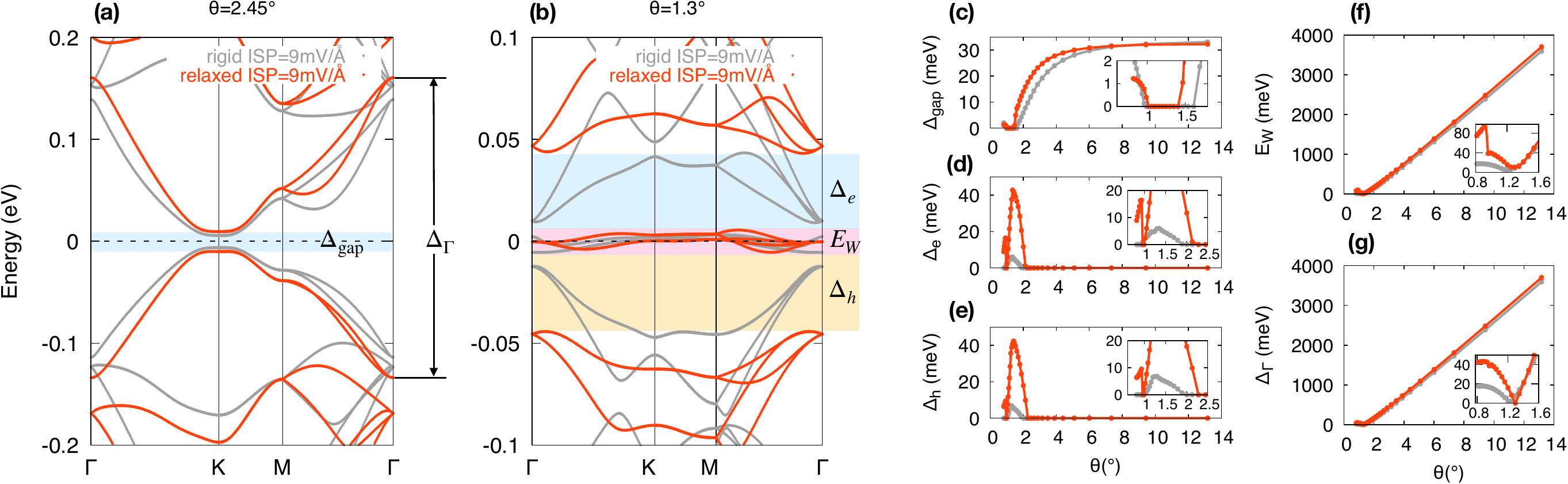}
 \caption{\label{fig6} {Band structures of rigid and relaxed TDBG at (a) $\theta = 2.54^{\circ}$ and (b) $\theta = 1.30^{\circ}$ calculated using the TB+ISP model. 
Dependence of (c-e) band gaps $\Delta_{gap}$, $\Delta_{e}$ and $\Delta_{h}$, respectively, (f) band width $E_W$ and (g) $\Gamma$-point gap $\Delta_{\Gamma}$ on twist angle $\theta$.
Definitions of these band-structure parameters are given in panels (a) and (b).
 }}	
 \end{figure*}

\textit{Flat bands in the tight-binding model with lattice relaxation effects.}--- 
In order to extend the scope of our models to twist angles in the range relevant to the flat-band physics, we perform the tight-binding model calculations. Our tight-binding model 
\footnote{We consider a tight-binding model Hamiltonian for carbon atom $p_z$ orbitals
$ H = \sum_{i \ne j} V_{ij} a^{\dag}_i a^{}_j$, 
where $a^{\dag}_i, a^{}_j$  are the creation and annihilation operators. The $\sigma$-type and $\pi$-type Slater-Koster parameters contribute to $V_{ij} = V_{pp\pi} \sin^2 \theta 
+ V_{pp\sigma} \cos^2 \theta$, where $\theta$ is the angle between the orbital axes and $\vec R_{ij} = \vec{R}_i - \vec{R}_j$ connects the two orbital centers \cite{Slater1954}. For $\theta=\pi/2$ ($\theta = 0$) which corresponds to the pair of atoms in the same layer (the pair of atoms on top of each other) $V_{ij} = V_{pp\pi}$ ($V_{ij} = V_{pp\sigma}$).
The Slater-Koster parameters depend on the distance $r$ between two orbitals as
$  V_{pp\pi} (r) = V^0_{p p \pi} e^{q_\pi (1 - r/a_\pi)} F_c(r)$ and
 $V_{p p \sigma} (r) = V^0_{p p \sigma} e^{q_\sigma (1- r/a_\sigma)} F_c (r)$, where $a_\pi$ is the first nearest neighbor distance in the plane and $V^0_{p p \pi} $ is the corresponding coupling value. In our work we assume $V^0_{p p \pi} = -2.81$~eV, which is larger than the conventional value of $-$2.7~eV \cite{Laissardiere2012}, in order to provide a better fit to our DFT calculations. The second nearest neighbour coupling of $0.1 \times V^0_{p p \pi}$ fixes the value of $q_{\pi} / a_{\pi}$. Here, $a_\sigma$ is the interlayer distance in AB-stacking BLG. In our relaxed structures the interlayer distance between the surface and inner layers is $3.364$~\AA\ and between two inner layers is $3.410$~\AA. Choosing the same exponential decay for both hopping parameters gives $q_{\sigma} / a_{\sigma} = q_{\pi} / a_{\pi}$.
$F_c$ is a smooth cutoff function that takes into account the distance between orbitals 
$ F_c(r) = (1+ e^{(r-r_c)/l_c})^{-1}$ \cite{Laissardiere2012}, 
where $l_c = 0.265$~\AA, $r_c = 2.5a = 6.165$~\AA. For $r \ll r_c$, $F_c(r) \simeq 1$ and for $r \gg r_c$, $F_c(r) \simeq 0$.
The on-site energy of $p_z$ orbitals is set to $\epsilon_i = -0.78$~eV to adjust the reference. 
All tight-binding parameters being defined, the band structures were calculated using WannierTools 
\cite{WU2017}. } 
 is based on the one described in Ref.~\cite{Laissardiere2012} with the ISP being taken into account as described above. Figure~\ref{fig3} evidences excellent agreement with the DFT results. Furthermore, we explicitly include lattice relaxation which becomes crucial at small twist angles using the methodology established by us previously \cite{Gargiulo2017}. Important details are summarized in Ref.~\footnote{
The lattice relaxation of TDBG models was performed in the classical force-field approach using the LAMMPS package \cite{LAMMPS}. The second generation REBO potential \cite{REBO} and the Kolmogorov-Crespi (KC) potential were used to describe the intra-layer and inter-layer interactions, respectively. 
The initial inter-layer distance was set as 3.35~\AA\ and the $D_3$ symmetry was preserved during the relaxation. The relaxation was performed using the FIRE algorithm \cite{FIRE} until the total force acting on each atom becomes less than 10$^{-6}$~eV/atom. Details of the relaxation results are presented in Ref.~\cite{SI}. 
}  
Figure~\ref{fig6}a shows the band structure of TDBG at twist angle $\theta = 2.45^\circ$, which is above the magic angle $\theta^\star$=1.3$^\circ$, for both rigidly twisted and relaxed models. 
In this case, the lattice relaxation increases the band gap $\Delta_{gap}$ produced by the ISP. Further decrease of the twist angle closes this band gap and results in a manifold of very narrow bands separated from the rest of bands by gaps $\Delta_e$ and  $\Delta_h$ (Fig.~\ref{fig6}b). Importantly, taking into account the relaxation effects increases these gaps dramatically, resulting in close-to-maximum values of {$\Delta_e = 37.6$}~meV and {$\Delta_h = 38.8$}~meV at $\theta$=1.3$^\circ$  (Fig.~\ref{fig6}b). At the same time, the width of the flat-band manifold {$E_W = 11$}~meV is achieved. Figures~\ref{fig6}c-g summarize the dependence of crucial band structure parameters $\Delta_{gap}$, $\Delta_e$, $\Delta_h$ and $E_W$ as well as the $\Gamma$-point gap $\Delta_\Gamma$ for the entire ensemble of investigated TDBG models with twist angles down to 0.8$^\circ$. The following picture of the twist-angle dependence emerges.
Firstly, as $\theta^\star$=1.3$^\circ$ is approached from above  ($\theta> \theta^\star$), the ISP-induced energy gap $\Delta_{gap}$, which is of order 30~meV outside of the magic angle region (Fig.~\ref{fig6}c), rapidly collapses to zero when the flattened bands are pre-formed and reopens only at  $\theta < 1^\circ$.
Secondly, the bandwidth of the flat-band manifold $E_W$ is dramatically suppressed at the magic angle $\theta^\star = 1.3^{\circ}$, where the bands become remarkably flat (Figs.~\ref{fig6}b,f).
Thirdly, within the same narrow region around $\theta^\star$, energy gaps $\Delta_e$ and $\Delta_h$ are maximized. The flat-band manifold is separated from the rest of bands within a twist angle region $1^\circ < \theta < 2^\circ$ (Figs.~\ref{fig6}d,e).
Yet the magic angle 1.3$^{\circ}$ is a well-defined quantity at which the band width is minimized, while the gap to excited states is close-to-the-maximum value of 37.6~meV (the maximum is offset towards 1.35$^{\circ}$ degrees at which the $\Delta_e$=42.7 meV). Figures~\ref{fig6}c-g also show
that the relaxation effects are pivotal at the magic angle, not only flattening the entire band manifold, but also increasing by an order of magnitude the gap to excited states (Fig.~\ref{fig6}b), making the TDBG half-bandwidth/bandgap ratio $14.7 \%$ as favorable as in the parent TBG superlattice \cite{Carr2019}. 

\textit{Origin of band flattening.}--- We can address the origin of the band flattening at the magic angle $\theta^\star = 1.3^{\circ}$ by constructing a minimalistic continuum model based on Refs.~\onlinecite{magic1,Guinea2012}. At small twist angles, the distinction between commensurate and incommensurate moir{\'e} structures is experimentally not relevant, and we can treat the physically important moir{\'e} branch $L(\theta) = a/2 \sin (\theta/2)$ as a continuum function.  
To construct the minimal TDBG Hamiltonian, we start from an effective model for the Bernal-stacked bilayer graphene \cite{CastroNeto2009} 
\begin{align}
\mathcal{H}_{\text{AB}} = 	
\begin{pmatrix}
v_F \, \boldsymbol \sigma \cdot \vec k & T_0
\\
T_0^{\dag}	& v_F \, \boldsymbol \sigma \cdot \vec k 
\end{pmatrix},
\ \ \ 
T_0 =
\begin{pmatrix}
\gamma_4 k &  \gamma_3  k^*
\\
\gamma_1 & \gamma_4 k 	
\end{pmatrix}, 	
\nonumber
\end{align}
\noindent
where $k = k_x + i k_y$. The terms $\gamma_3$ and $\gamma_4$ represent trigonal warping and particle-hole asymmetry.  When two graphene bilayers are twisted to form TDBG, the inner graphene layers (2nd and 3rd) are coupled through the twist-induced interlayer coupling $T_{\theta}(\vec r)$. 
To treat both $T_0$ and $T_{\theta}$ consistently, for now we neglect the $\gamma_3$ and $\gamma_4$ terms, which gives
\begin{align}
T_{\theta}(\vec r)  = \sum_{n =1,2,3}
T_n \, e^{- i \vec q_n \vec r}, 
\end{align}
\no
with a moir{\'e} three-fold star of $\vec q_i$,  $|\vec q_j| = 2 k_D \sin \nicefrac{\theta}{2}$, equirotated by  $\phi = 2 \pi/3$, and
\begin{align}
T_n  = e^{- i \boldsymbol{\mathcal{G}}^{(n)}_{\theta} \vec d} \  
\hat \Omega_{\phi}^{n-1}  
\begin{pmatrix}
w_{AA} & w_{AB} 
\\
w_{AB} & w_{AA}  	
\end{pmatrix}
\hat \Omega_{\phi}^{1-n},   	
\end{align}
\noindent
with $\hat \Omega_{\phi} = \cos \phi \, \sigma_x - \sin \phi \, \sigma_y$. Here, $\boldsymbol{\mathcal{G}}^{(0)}_{\theta} = 0$, $\boldsymbol{\mathcal{G}}^{(1)}_{\theta} = \vec q_2 - \vec q_1$, $\boldsymbol{\mathcal{G}}^{(2)}_{\theta} = \vec q_3 - \vec q_1$ are the moir{\'e} reciprocal cell vectors  and $\vec d$ is the relative displacement of one bilayer with respect to another one. The effective TDBG Hamiltonian thus reads
\begin{align}
\mathcal{H} = 	
\begin{pmatrix}
\mathcal{H}_{1} & T_0 & 0 & 0
\\
T_0^{\dag}	& \mathcal{H}_{2} & T^{}_{\theta}(\vec r) & 0 \\
0 & T^{\dag}_{\theta}(\vec r) & \mathcal{H}_{3} & T_0 
\\
0 & 0 & T^{\dag}_0 & \mathcal{H}_{4} 
\end{pmatrix}
\label{Ham TDBG} ,
\end{align}
\noindent
where  $\mathcal{H}_{1,2} = - i \boldsymbol \sigma_{-\theta/2}\boldsymbol{\nabla} + \Delta_{1,2}$ and $\mathcal{H}_{3,4} = - i \boldsymbol \sigma_{+\theta/2}\boldsymbol{\nabla} + \Delta_{3,4}$, where $\Delta_{i}$
is electric potential on the i$^{\text{th}}$ layer.

\begin{figure}[t]
 \includegraphics[width=  \columnwidth]{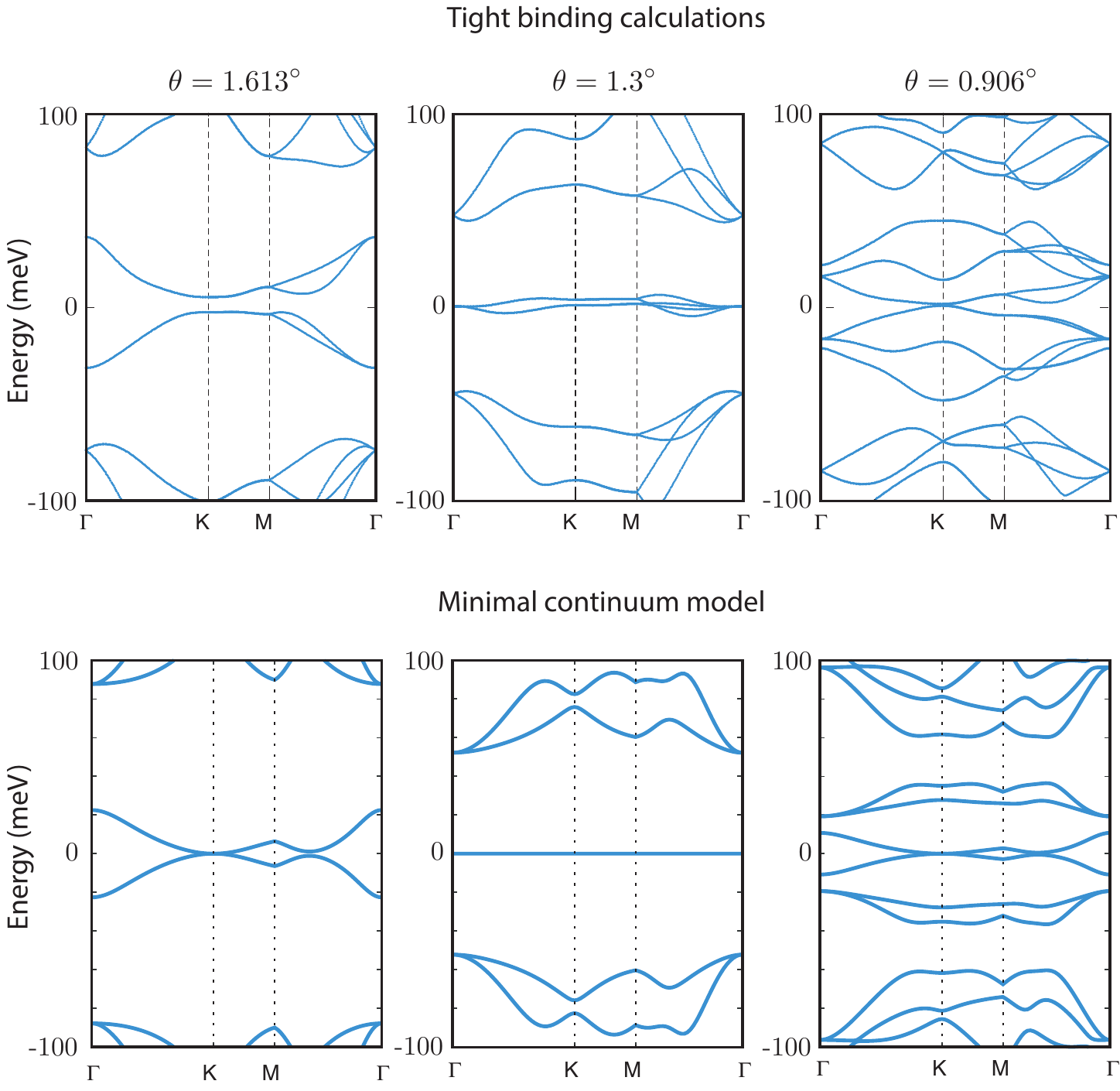}
 \caption{\label{fig_cont} 
 Band structures of TDBG at and in the vicinity of magic angle  $\theta^\star = 1.3^{\circ}$ obtained from the tight-binding model with lattice relaxations (upper row) and the minimal continuum model (lower row). For the continuum model, the following parameters have been used: $\hbar v_F k_D =9.78$~eV, $w_{AB} = 130$~meV, $w_{AA} = 0$, $\Delta_i = 0$.    }
 \end{figure}

To understand the origin of band flattening at the magic angle, we further construct the minimal model reproducing perfectly flat bands. For this, we for a moment neglect $\Delta_i$ and switch off $w_{AA}$, which is a natural consequence of lattice relaxation at small twist angles \cite{magic1}. 
To be self-consistent in $T_0$ and $T_\theta$, we set $\gamma_1 = 3 w_{AB}$,   thus condition $T_{\theta = 0} (\vec d_{AB}) = T_0$ is satisfied. In this limit, the minimal TDBG Hamiltonian \eqref{Ham TDBG} has particle-hole and chiral symmetries.  We see that upon these imposed conditions the bands become absolutely flat at the magic angle (Fig.~\ref{fig_cont}). Similar to the TBG case, this absolutely flat band becomes highly dispersive both below and above the magic angle, making the value of the magic angle rigorously defined. One can show that the TDBG Hamiltonian maps directly onto the TBG case in the chirally-symmetric limit. Indeed, Hamiltonian \eqref{Ham TDBG} can be rewritten as
\begin{align}
\mathcal{H} = 
\begin{pmatrix}
0 & D^{\dag}(\vec r) 
\\
D^{}(\vec r) & 0	
\end{pmatrix}, 
\ \ \ 
D^{}(\vec r)	= 
\begin{pmatrix}
- 2 i \bar \partial & \alpha \hat A_1 (\vec r) 
\\
\alpha \hat A_2 (\vec r) & - 2 i \bar \partial
\end{pmatrix}, 
\end{align}

\no
which reminds the TKV model for TBG \cite{magic1}. One can further show that this Hamiltonian maps onto two TBG Hamiltonians~\cite{SI}, where the band flatness comes mathematically from the flatness of the lowest Landau level in the quantum Hall effect on torus \cite{magic1,Liu2019-Landau}.   
Therefore, the band flatness in the magic-angle TDBG is of  the same topological nature. More realistic description of the flat bands in TDBGs at magic angle $\theta^\star$ would require including the interplay between trigonal warping, particle-hole asymmetries and the ISP fields as a perturbation around the perfectly flat bands, but the well-defined magic angle would be preserved.

To our surprise, the introduced minimalistic model is qualitatively consistent with our atomistic calculations (Fig.~\ref{fig_cont}), including the reasonable values of band gaps both at the magic angle and in its vicinity. We stress that the effect of the ISP field is more important at higher angles, opening a gap between otherwise touching parabolic bands. On contrary, at the magic angle, the lattice relaxation effects and particle-hole asymmetries suppress the ISP effect (Fig.~\ref{fig6}b), so we neglect this intrinsic polarization in the continuum model results shown in Fig.~\ref{fig_cont}.   
The effects of lattice relaxation on the band structure of TDBG are more complex than in the case of TBG, and cannot be qualitatively accounted for in the continuum model by simple reduction of $w_{AA}$. For example, taking  trial parameters $\gamma_3$, $\gamma_4$ expected from the BLG case and estimative $w_{AA} \approx 0.8 \, w_{AB}$ does not reproduce the band gap and band width at the magic angle, when compared to the tight-binding values. We anticipate the values of $\gamma_3$ and $\gamma_4$ to be significantly renormalized by the lattice relaxation effects, which we will address in a further study.  

\textit{Conclusions.}---With the help of atomistic calculations we show that twisted double bilayer graphene has a well-defined magic angle, at which it hosts isolated flat bands, gapped out from higher excited states by dramatic 38~meV. In terms of the relative flatness (bandwidth/bandgap ratio), the TDBG at the magic angle 1.3$^{\circ}$ is 
very close to the parent TBG heterostructure.  A direct algebraic mapping to the TBG case can be found in the continuum setting if particle-hole asymmetries are neglected. 
A surprising novel detail revealed by the DFT calculations is the intrinsic symmetric polarization of the TDBG layers, which has not been reported previously. This internal effect modifies the magic-angle band dispersion. Previously, it was considered that the bands in TDBG can be made relatively flat only by applying external electric fields \cite{Liu2019,Koshino2019}. We however report that the bands in TDBG are naturally flat at the magic angle 1.3$^{\circ}$ due to significant intrinsic polarization and lattice relaxation effects. Further application of electric fields provides a control mechanism over the bandwidth and flat-band structure, important for the fine-tuning with respect to the electron-electron interaction scale \cite{Liu2019,Cao2019}.
Flat bands can promote exotic correlated states, such as unconventional superconductivity, fractional quantum Hall effect and ferromagnetism in the flat bands.  These anticipated phenomena in TDBG remain to be addressed.

\textit{Acknowledgements}. This work was supported by NCCR Marvel and the Swiss National Science Foundation (grant No.~P2ELP2\_175278). We thank Fernando Gargiulo for assistance.


%

\end{document}